\documentstyle[Qqa4lart]{article}

\input Tcilatex

\begin{document}

\begin{center}
{\bf Possibility of Microturbulence Diagnostics in a Magnetically Confined
Plasma Using Multiple Scattering Effects\\[20mm]}

{\rm E.S.Kovalenko} and {\rm N.A.Zabotin}{\bf \\[20mm]}
\end{center}
\begin{abstract}
The idea of new diagnostics method for the small-scale irregular structures
of magnetically confined plasma is suggested in the present paper. The
method can be based on measurements of intensity attenuation of the normal
sounding waves. Anomalous attenuation arises due to multiple scattering
effects investigated earlier for ionospheric radio propagation. It has been
shown that multiple scattering regime can realize in a tokamak plasma.
Calculations of normal sounding wave anomalous attenuation in a tokamak
plasma have been carried out. This quantity is large enough to be registered
experimentally.
\end{abstract}
\section{Introduction}

Anomalously large level of energy and particle transport is one of the main
problems in the magnetic confinement fusion research. The transport is
thought to be enhanced by small scale plasma turbulence [1-3]. Therefore,
the determination of the microturbulent fluctuations properties is necessary
for understanding and improvement of plasma confinement.

At present, there are several basic methods of microturbulence diagnostics:
Langmuir probes, heavy ion beam probes, scattering, beam emission
spectroscopy, electron cyclotron emission, reflectometry and some others.
Each of these techniques has limited field of application, merits and
demerits [4]. Joint usage of different methods allows to obtain more
valuable and more accurate information about microturbulence properties. The
application of the existing methods at large devices of the future (such as
ITER) requires additional studies and probably some of them will not be
possible or will become more difficult there. Therefore, the development of
new diagnostic methods is an actual problem.

In this paper it is suggested the idea of using of electromagnetic wave
multiple scattering effects for diagnostics of spatial spectrum of small
scale electron density fluctuations in a magnetically confined plasma.
Unlike existing scattering technics, instead of the scattered field
registration, it is suggested to measure the power of signal reflected from
plasma. The theoretical basis of the idea has been taken from the works
dealing with the application of the multiple scattering theory to the
ionosphere radio wave propagation. The attenuation of the vertical sounding
signal is one of the consequences of the theory. This phenomenon is well
known and it has been observed in a number of ionospheric experiments [5-7].
However, ionosphere parameters, properties of random irregularities and
sounding frequencies strongly differ from those in laboratory plasma. For
instance, maximum average electron density in the ionosphere is $10^6$ $%
cm^{-3}$ and in the magnetically confined plasma is $10^{13}-10^{14}\
cm^{-3} $. Typical sounding signal frequency for reflection case is about $%
10\ MHz$ in the ionosphere and $10-100$ $GHz$ in tokamak plasma. Small scale
ionospheric irregularities of importance for the scattering process have
size of $1-10$ $km$ across the magnetic field and in the tokamak plasma the
irregularities are of $1-5$ $cm$ in diameter. This strong difference of
parameters demands separate study for the case of high-temperature
magnetically confined plasma. Such study is presented in this paper.

The analysis and numerical estimates are carried out for the large tokamak
(with minor radius $a\geq 1$ $m$), however, the suggested idea can also be
realized for the next generation devices, such as NSTX [8].

We are starting the paper with the description of the suggested experiment
scheme, presentation of the electron density irregularity spatial spectrum
model and estimation of plasma optical depth. Then, the radiative transfer
equation in a randomly irregular magnetized plasma and its approximate
analytical solution are presented. The next section is devoted to the
analysis of the applicability of this theory to the tokamak plasma. Then,
the results of numerical estimations of the anomalous attenuation effect are
presented. Finally, the obtained results are discussed and the conclusions
are presented.

\section{Multiple scattering regime in a tokamak plasma}

Scheme of the experiment for density fluctuation study is shown in Fig. 1.
It is suggested to measure the intensity of normally reflected signal. If
the scattering is multiple, than, according to the existing theory [9], in
the case of the normal sounding this can cause considerable attenuation of
the reflected signal. The attenuation value is proposed to be used for
irregularities study. To find out the conditions, in which the scattering is
multiple, it is necessary to carry out the estimation of plasma slab optical
thickness. The optical thickness $L$ for scattering process is determined by
the expression 
\begin{equation}
\label{Eq1}L=\int \sigma _0dS\text{,} 
\end{equation}
where $\sigma _0$ is full scattering cross-section of a unit volume, $dS$ is
the element of nonperturbed ray trajectory. The value of $\sigma _0$ is
determined by integration of the differential scattering cross-section $%
\sigma $ over full solid angle 
\begin{equation}
\label{Eq2}\sigma _0=\int\limits_{4\pi }\sigma d\Omega \text{ .} 
\end{equation}

The calculation will be carried out in the isotropic plasma approximation.
It is implied using the expression of the differential scattering
cross-section for isotropic plasma. Unlike of the isotropic plasma case, the
expression for differential cross-section for magnetized plasma contains
dimentionless multiplier (so called ''geometrical factor''), depending on
the polarization of incident and scattered waves [10]. If the wave frequency
is not close to some plasma resonance, then the geometrical factor is about
a unit. We will neglect of this cross-section dependence on the polarization
and set the geometrical factor equal to unit. Also we will use the
refractive index for isotropic plasma.

Utilization of this simplifications is justified by calculation results for
the ionosphere. Calculation of the ionosphere optical depth does not lead to
considerable quantitative difference from the same calculation in the
isotropic plasma approximation [9].

Although tokamak plasma temperature is high ($T\sim 10^8$ $K$), the electron
thermal motion in our problem does not have considerable influence on
radiation propagation. It is bound up with the fact that in the case of
normal sounding we are interested in waves propagating nearly
perpendicularly to the magnetic field. In this case temperature correction
for the refractive index is exponentially small [11] and one can use the
refractive index for a cold plasma ($n=1-v$,$\ v=\omega ^2/\omega _e^2$, $%
\omega _e$ - plasma frequency, $\omega =2\pi f$, $f$ - wave frequency).

The differential scattering cross-section in isotropic plasma takes the form
[12] 
\begin{equation}
\label{Eq3}\sigma \left( \alpha _0,\beta _0,\alpha ,\beta \right) =\frac \pi
2k_0^4v^2F\left[ \vec k^{\prime }\left( \alpha ,\beta \right) -\vec k\left(
\alpha _0,\beta _0\right) \right] \text{ ,} 
\end{equation}
where $k_0=\omega /c$, $F\left( \vec k\right) $ is density fluctuation
spatial spectrum, $\alpha _0,\beta _0,\alpha ,\beta $ - polar and azimuthal
angles of wave vectors $\vec k$ and $\vec k^{\prime }$ of incident and
scattered waves, respectively.

For numerical calculation of optical depth it is necessary to concretize the
model of irregularity spectrum being based on existing experiment
information. According to experimental data, the irregularities in tokamak
plasma are strongly stretched along the magnetic field: $l_{\mid \mid }\sim
100-1000\ cm$, $l_{\perp }\sim 1-5\ cm$, where $l_{\mid \mid }$ and $%
l_{\perp }$ are typical irreglarity sizes in the parallel and perpendicular
to the magnetic field directions. Since longitudinal sizes exceed transverse
ones by $100-1000$ times, we can use for our estimates the approximation of
infinitly stretched irregularities with spatial spectrum 
\begin{equation}
\label{Eq4}F\left( \vec k\right) =C_A\left( 1+\frac{\kappa _{\bot }^2}{%
\kappa _{0\bot }^2}\right) ^{-\nu /2}\delta \left( \kappa _{_{\parallel
}}\right) \text{ ,} 
\end{equation}
where $C_A$ is normalizing constant, $\kappa _{0\perp }=2\pi /l_{0\perp }$, $%
l_{0\perp }$ is external irregularity scale length, $\kappa _{\perp }$ and $%
\kappa _{\mid \mid }$ are transverse and longitudinal to the magnetic field
components of the irregularity spatial harmonic, $\nu $ is spectrum index
and $\delta (x)$ is delta-function. Spectrum (\ref{Eq4}) dependence on $%
\kappa _{\perp }$ for $\kappa _{\perp }>>\kappa _{0\perp }$ takes form $%
F\sim \kappa _{\perp }^{-\nu }$ what is consistent with the existing
experimental data for $\nu $ from $2$ to $3.5$ [4,13].

For the spectrum normalization a certain value of the relative
irregularities level in a some scale $R$ can be used. The most natural
analog of this physical value in the locally homogeneous random field theory
is the structural function [12] 
\begin{equation}
\label{Eq5}D\left( \vec R\right) =\left\langle \left[ \frac{\delta n_e}{n_e}%
\left( \vec r+\vec R\right) -\frac{\delta n_e}{n_e}\left( \vec r\right)
\right] ^2\right\rangle \text{ ,} 
\end{equation}
where $\frac{\delta n_e}{n_e}\ $is relative electron density perturbation, $%
\left\langle {}\right\rangle $ means ensemble average. To determine the
normalizing constant $C_A$ we will normalize the spectrum (\ref{Eq4}),
following to the method of [9], by the value of structural function (\ref
{Eq5}), choosing irregularity scale length $R$ in transverse to the magnetic
field direction being corresponded to the interested spectrum interval. An
important property of the structural function is that perturbations $\frac{%
\delta n_e}{n_e}$ of large spatial scale lengths (with typical size $l>>R$)
do not have influence on it. The structural function is connected with
spatial spectrum by the following expression [12] 
\begin{equation}
\label{Eq6}D\left( \vec R\right) =2\int F\left( \vec \kappa \right) \left(
1-\cos \vec \kappa \vec R\right) d^3\kappa \text{ .} 
\end{equation}

Thus, setting relative density perturbation $\frac{\delta n_e}{n_e}=\delta
_R $ in a certain scale $R$, assuming $D\left( \vec R\right) \equiv \delta
_R^2$ and using then formula (\ref{Eq6}) we determine the normalizing
constant: 
\begin{equation}
\label{Eq7}C_A=\delta _R^2\frac{\Gamma (\nu /2)}{2\pi \kappa _{0\perp }^2}%
\left[ \Gamma \left( \frac{\nu -2}2\right) -2\left( \frac{R\kappa _{0\bot }}%
2\right) ^{\frac{\nu -2}2}K_{\frac{\nu -2}2}(R\kappa _{0\bot })\right] ^{-1}%
\text{ ,} 
\end{equation}
where $\Gamma (x)$ is gamma-function, $K(z)$ is McDonald function [14].

Numerical calculation of the optical depth for the ray trajectory with
coinciding incidend and reflected ray paths (see Fig. 1, where, however, the
incident and reflected rays are drown separately for clearness) was carried
out for the linear regular density profile with $n_e=10^{14}\ cm^{-3}$ at
distance $100\ cm$ from the slab boundary (see Fig. 2). The following values
of spectrum parameters were chosen: external irregularity scale length $%
l_{0\perp }=5\ cm$, spectrum index $\nu =2.5$, irregularity level $\delta
_R=1;1.5;2\%$, normalization scale $R=1\ cm$. The magnetic field direction
was chosen perpendicular to the ray path. The value of magnetic field is of
no importance in used approximation, but its direction determines the
irregularity orientation. The calculation results are shown in Fig. 3 in the
form of dependence of the optical depth $L$ on the sounding wave frequency $%
f $. Chosen frequency interval corresponds to wave penetration depth from $%
z=50\ cm$ to $z=100\ cm$. The obtained results show that in the chosen
frequency (or reflection level) band the optical depth is considerably more
than a unit (unit optical depth corresponds to $L=4.3$ $dB$) for relative
irregularity level $\frac{\delta n_e}{n_e}\geq 1\%$.

Thus, for parameters characterised of tokamak irregularities and plasma, the
realization of the multiple scattering mode is possible.

\section{Radiation transfer in a randomly irregular magnetized plasma}

In the considered case of normal sounding, the rays situated near the normal
ray trajectory give the basic contribution to the reflected signal power.
That is why we will assume the plasma layer to be plane stratified. As it
has been shown in [15,16], radiation energy transfer with multiple
scattering effects accounting for the case of total internal reflection from
a plane stratified layer of randomly irregular plasma can be described by
the equation of radiation energy balance (REB) in ray tubes. This equation
is written in terms of the invariant ray variables (coordinates). The latter
ones permit to take into account naturally of regular refraction and give
the most simple form to the equation.

The invariant ray variables are introduced by setting the basic plane out of
the layer and parallel to it. Let us introduce Cartesian orthogonal
coordinates $(x,y,z)$ with $z$-axis being directed along the plasma density
gradient. Then $XOY$ plane can be considered as the basic plane. The plasma
occupies the region $z>h_0$ (see Fig. 4). The coordinates $\vec \rho =(x,y)$
of intersection point of a ray trajectory going out of the layer with the
basic plane as well as ray polar $\theta $ and azimuthal $\varphi $ arrival
angles in this point completely determine ray trajectory within the plasma
layer and outside of it. In this meaning they are called ''invariant''. The
equation has the following form 
\begin{equation}
\label{Eq8}\frac d{dz}P(z,\vec \rho ,\omega )=\int Q(z,\omega ,\omega
^{\prime })\left\{ P(z,\vec \rho -\vec \Phi (z,\omega ^{\prime },\omega
),\omega ^{\prime })-P(z,\vec \rho ,\omega )\right\} d\omega \,\,\,\text{,} 
\end{equation}
where $\omega =\{\theta ,\varphi \}$; $d\omega =d\theta d\varphi $; $P$ -
radiation energy flux density in a unit solid angle in direction determined
by angles $\theta $, $\varphi $, at the point $\vec \rho $ of basic plane; 
$$
Q(z,\omega ,\omega ^{\prime })=\sigma (\omega ,\omega ^{\prime
})C^{-1}(z,\omega )\sin \theta ^{\prime }\left| \frac{d\Omega }{d\Omega
^{\prime }}\right| \text{ ,} 
$$
$\sigma (\omega ,\omega ^{\prime })$ is scattering differential
cross-section, $C^{-1}(z,\omega )$ is cosine of angle between ray trajectory
and $z$-axis at level $z$, $\left| \frac{d\Omega }{d\Omega ^{\prime }}%
\right| $ is Jacobian of transition from current wave vector angles to
invariant ones, $\vec \Phi (z,\omega ^{\prime },\omega )$ is vector
connecting points of intersection with basic plane of two ray trajectories
determined by invariant angles $\omega $ and $\omega ^{\prime }$ under the
condition that trajectories intersect each other at level $z$. Using of the
invariant ray coordinates allows one to introduce the small angle scattering
approximation in the invariant ray coordinates [9]. This approximation is
valid if the most probable difference of invariant angles in each scattering
act is small. It must be noticed that the applicability field of this
approximation is somewhat more wide then that of the ordinary small angle
scattering approximation. In particular, when the scattering occures near
reflection level, small difference of the invariant angles can correspond to
considerable difference of wave vector orientation angles. This
approximation allows one to obtain an analytical solution of the equation
(8). The solution consists of two terms. The first term gives the basic
radiation energy flux [9] 
\begin{equation}
\label{Eq9}
\begin{array}{c}
\tilde P(z,\vec \rho ,\omega )=\frac 1{\left( 2\pi \right) ^2}\dint
d^2qP_0(\vec q,\omega )\cdot \\ 
\cdot \exp \left\{ i\vec q\vec \rho +\int\limits_0^zdz^{\prime }\int d\omega
^{\prime }\,Q(z^{\prime },\omega ,\omega ^{\prime })\left[ e^{-i\vec q\vec
\Phi (z^{\prime };\omega ,\omega ^{\prime })}-1\right] \right\} \,\,\text{,} 
\end{array}
\end{equation}
where $P_0(\vec q,\omega )$ is the Fourier transform of the energy flux
spatial-angular distribution $P_0(\vec \rho ,\omega )$ of the radiation
reflected from the layer in absence of irregularities. This undisturbed flux
is determined by the source directivity diagram and the regular layer
parameters. The second term (not shown in (\ref{Eq9})) has the sense of
difference between approximate and exact solutions of the equation (\ref{Eq8}%
). It may be shown using asymptotic estimates that under considered
approximation the second term is small.

\section{Applicability of the radiation transfer theory for tokamak plasma}

In the next section the outlined theory will be applied to calculation of
the normal sounding signal attenuation in a tokamak plasma layer. But before
that, the applicability analysis of used approximations has to be carried
out. First of all, for the transfer theory utilizing it is necessary to
clarify the applicability of the geometrical optics approximation for the
average field. The radiation wave length ($\lambda \sim 0.3\ cm$ for $f\sim
100\ GHz$) must be much less then the average density regular distribution
scale length. If the density profile is sufficiently smooth, then this scale
length is about the tokamak minor radius ($a\sim 1\ m$ for large devices).
So, the geometrical optics approximation is valid in this case.

The next assumption to be verified is the validity of the small angle
scattering approximation in the invariant ray coordinates. The frequency
band of interest is $60-90\ GHz$, what corresponds to wave length band of $%
0.3-0.5\ cm$, but the minimum irregularity scale length is $1$ $cm$, and it
is at least two times larger than the wave length. It means that in the
entry region of the plasma layer the usual small angle scattering
approximation is valid. In the plasma layer depth, near the wave reflection
level, as it was mentioned in the previous section, there exist additional
reasons for using the small angle scattering approximation in the invariant
ray coordinates.

\section{Numerical culculation of the reflected wave attenuation uder normal
sounding of tokamak plasma layer}

For the calculation of the signal attenuation due to multiple scattering we
use the solution (\ref{Eq9}) of the equation (\ref{Eq8}) in the small angle
scattering in the invariant ray coordinates approximation. We assume the
antenna to have small sizes and wide directivity diagram. Hence, this source
may be approximately considered as point. Let the antenna be situated in the
coordinate center, point $O$ (see Fig. 4), at the distance $h_0=10\ cm$ from
the layer boundary.

Then, we take the function $P_0$ in the form 
\begin{equation}
\label{Eq10}P_0(\vec \rho ,\theta ,\varphi )=\tilde P_0(\vec \rho )\delta
\left[ -\cos \theta +\cos \theta _0(\vec \rho )\right] \delta \left[ \varphi
-\varphi _0(\vec \rho )\right] \text{ ,} 
\end{equation}
where $\theta _0(\vec \rho )$ and $\varphi _0(\vec \rho )$ are angle
coordinates of the ray coming to the point $\vec \rho $ when neglecting the
scattering.

The calculation is carried out for the same linear density layer (Fig. 2) of
a cold isotropic plasma and for the same frequency interval what have been
used in section 2 for the optical depth estimates. The function $\vec \Phi $
for the plane isotropic plasma layer can be obtained in analytical form 
\begin{equation}
\label{Eq11}
\begin{array}{c}
\Phi _x(v,\theta ,\varphi ,\theta ^{\prime },\varphi ^{\prime })=f(\theta
^{\prime })\cos \varphi ^{\prime }-f(\theta )\cos \varphi 
\text{ ,} \\ \Phi _y(v,\theta ,\varphi ,\theta ^{\prime },\varphi ^{\prime
})=f(\theta ^{\prime })\sin \varphi ^{\prime }-f(\theta )\sin \varphi \,%
\text{ ,}\, 
\end{array}
\end{equation}
where%
$$
f(\theta )=2H\sin \theta \left( \cos \theta +\sqrt{n^2-\sin ^2\theta }%
\right) +h_0\tan \theta \text{ , } 
$$

$$
H=dv/dz\text{ .} 
$$

The intensity of normally reflected signal is obtained using formula (\ref
{Eq9}), after substituting of $\vec \rho =0$ and integration over angle
variables. In view of the integrand complicity, the calculation is carried
out numerically. The numerical results for various irregularity spectrum
parameters are shown in Fig. 5 in a form of dependence of the signal
attenuation on frequency.

The attenuation for three different irregularity levels $\delta _R=1;1.5;2\%$
and $l_0=10\ cm$, $\nu =2.5$ is illustrated in Fig. 5(a). The first, quite
natural conclusion, is that the attenuation increases with the fluctuation
amplitude. The results obtained for $l_0=3,5,10\ cm$, $\nu =2.5$ and $\delta
_R=1.5\%$ are presented in Fig. 5(b). The attenuation slightly increases
with the external irregularity scale length. Figure 5(c) shows the results
obtaned for $\nu =2.5,2.75,3$, $l_0=10\ cm$ and $\delta _R=1.5\%$. One can
see that the attenuation also grows with incsease of the spectrum index $\nu 
$. Finally, all Figs. 5 (a)-(c) show the attenuation growth with frequency
increase. In the chosen frequency band ($60-90$ $GHz$) total attenuation
variation is $2-4$ $dB$.

The main feature of the presented results, of importance for the present
paper basic topic, is that the signal attenuation caused by scattering
amounts of $3-7$ $dB$ and can be measured in experiment.

\section{Conclusion}

The paper considered the problem of the sounding electromagnetic wave
propagation in a magnetically confined plasma with accounting of multiple
scattering effects. It was shown that, for typical tokamak plasma and
irregularity parameters, the multiple scattering regime can take place. The
anomalous attenuation of the normal sounding signal is one of the
consequences of this fact. The numerical calculations of the anomalous
attenuation were carried out. It was shown that the attenuation value is
sufficiently large to be registered by experimental facilities. The
attenuation dependences on signal frequency and irregularity spatial
spectrum parameters were obtained. Since the anomalous attenuation depends
on the spectrum parameters, its measuring can be used for stating and
solving of the inverse problem. Thus the aim of the irregularity
characteristics determination using observations of the attenuation can be
reached. Utilization of this method can broaden the possibilities of the
existing microturbulence diagnostic methods.

\section{References}

1. E.J. Doyle, K.H. Burrell, T.N. Carlstrom et al., Proc. of the 16th IAEA
Fusion Energy Conf., Montreal, Canada, 1996 (International Atomic Energy
Agency, Vienna, to be published).

2. R. L. Hickok, P.M. Schoch, T. P. Crowley et al., Nucl. Fusion Supplement,
1229 (1991).

3. C.L. Rettig, W.A. Peebles, K.H. Burrel, R.J. La Haye, E.J. Doyle, R.J.
Groebner and N.C. Luhmann, Jr., Phys. Fluids B 5, 2428 (1993).

4. N.Bretz. Rev. Sci. Instrum., 68, 2927 (1997).

5. V.V. Vyaznikov, V.V. Vaskov, Yu.V. Gruzdev, Geomagnetism and Aeronomy 18,
45 (1978), in russian.

6. E. Mjolhus, J. Geophys. Res. 90, 4269 (1985).

7. C.S.G.K. Setty, A.R. Jain, Canadian J. Phys. 48, 653 (1970).

8. http://www-local.pppl.gov/nstxhome/nstx/

9. N.A.Zabotin, Thesis... Dokt. of Phys.-Math. Sciences, RSU, Rostov-on-Don,
Russia, 1994, in russian.

10. Electrodinamics of Plasma, edited by A.I.Akhiezer, Moscow, ''Nauka'',
1974, in russian.

11. V.L.Ginzburg, Propagation of Electromagnetic Waves in Plasma, Moscow,
''Nauka'', 1967, in russian.

12. S.M. Rytov, Yu.A. Kravtsov, B.I. Tatarskii, Introduction to Statistical
Radiophysics, Part II, Random Fields, Moscow, ''Nauka'', 1978, in russian.

13. R.D. Durst, R.G. Fonck, G. Cosby, H. Evensen, S.F. Paul, Rev. Sci.
Instrum., 63, 4907 (1992).

14. Handbook of Mathematical Functions, edited by M. Abramovitz and I.A.
Stegun, Moscow ''Nauka'', 1979, in russian.

15. N.A. Zabotin, Izvestiya Vysshich Uchebnykh Zavedenii, Radiofizika, 36,
1075 (1993), in russian.

16. A.G. Bronin, N.A. Zabotin, Izvestiya Vysshich Uchebnykh Zavedenii,
Radiofizika, 36, 1163 (1993), in russian.

\end{document}